\documentclass[useAMS,usenatbib]{mn2e}
\usepackage{amsmath}
\usepackage{graphics}

\title[On the radio properties of HLX-1]{On the radio properties of the intermediate-mass black hole candidate ESO 243-49 HLX-1}
\author[D. Cseh et al.]{D. Cseh$^{1}$\thanks{E-mail:
d.cseh@astro.ru.nl}, N. A. Webb$^{2}$, O. Godet$^{2}$, D. Barret$^{2}$, S. Corbel$^{3}$, M. Coriat$^{4}$, 
\newauthor H. Falcke$^{1}$, S. A. Farrell$^{5}$, E. K\"{o}rding$^{1}$, E. Lenc$^{5,6}$, J. M. Wrobel$^{7}$ 
\\
$^{1}$Department of Astrophysics/IMAPP, Radboud University Nijmegen, P.O. Box 9010, 6500 GL Nijmegen, The Netherlands\\
$^{2}$IRAP, Universit\'e de Toulouse, UPS, 9 Avenue du colonel Roche, F-31028 Toulouse Cedex 4, France;\\ CNRS, UMR5277, F-31028 Toulouse, France\\
$^{3}$Laboratoire AIM (CEA/IRFU-CNRS/INSU-Universit\'{e} Paris Diderot), CEA/DSM/IRFU/SAp, F-91191 Gif-sur-Yvette, France\\
$^{4}$Astrophysics, Cosmology and Gravity Centre, Department of Astronomy, University of Cape Town,\\ Private Bag X3, Rondebosch 7701, South Africa\\
$^{5}$Sydney Institute for Astronomy, School of Physics, The University of Sydney, NSW 2006, Australia\\
$^{6}$ARC Centre of Excellence for All-sky Astrophysics (CAASTRO)\\
$^{7}$National Radio Astronomy Observatory, Socorro, NM 87801, USA\\
}
\begin{document}
\date{Draft}
\pagerange{\pageref{firstpage}--\pageref{lastpage}} \pubyear{2014}
\maketitle
\label{firstpage}

\begin{abstract}
We present follow-up radio observations of ESO 243-49 HLX-1 from 2012 using the Australia Telescope Compact Array (ATCA) and the Karl G. Jansky Very Large Array (VLA). We report the detection of radio emission at the location of HLX-1 during its hard X-ray state using the ATCA. Assuming that the `Fundamental Plane' of accreting black holes is applicable, we provide an independent estimate of the black hole mass of $M_{\rm{BH}}\leq2.8^{+7.5}_{-2.1} \times 10^{6}$~M$_{\odot}$ at 90\% confidence. However, we argue that the detected radio emission is likely to be Doppler-boosted and our mass estimate is an upper limit. We discuss other possible origins of the radio emission such as being due to a radio nebula, star formation, or later interaction of the flares with the large-scale environment. None of these were found adequate. The VLA observations were carried out during the X-ray outburst. However, no new radio flare was detected, possibly due to a sparse time sampling. The deepest, combined VLA data suggests a variable radio source and we briefly discuss the properties of the previously detected flares and compare them with microquasars and active galactic nuclei.

\end{abstract}


\section{Introduction}

Identifying intermediate-mass black holes (IMBHs; $M_{\rm{BH}}=10^2-10^5$~M$_{\odot}$) or their hosts is not straightforward \citep{Miller:2004uq}. The observational examples that are closest to this IMBH mass range are often termed as massive black holes (MBHs) and often considered to be in a range of $10^5-10^7$~M$_{\odot}$ \footnote{The term IMBH is used in the ULX community while the term MBH is used in the AGN community and these mass ranges are blurred together in the $10^4-10^6$~M$_{\odot}$ range. However, the low end of the IMBH range is physically motivated as BHs above 100~M$_{\odot}$ are hard to form from a single star. Also, one may want to avoid the usually considered lower range of supermassive BHs of $10^6$~M$_{\odot}$.}. Among these MBHs, NGC 4395 has the lowest mass black hole (BH) of $M_{\rm{BH}}=(3.6\pm1.1)\times10^5$~M$_{\odot}$ \citep{Peterson:2005uq}, that is obtained from reverberation mapping. There are several other examples of MBHs with mass estimates either based on their accretion properties \citep[e.g.][]{Reines:2011uq,Kamizasa:2012oq,Secrest:2012kx,Nyland:2012zr,Reines:2014ly} or are virial mass estimates \citep[e.g.][and references therein]{Greene:2004fk,Barth:2004qf,Seth:2010nx,Schramm:2013vn,Reines:2013ys}. The majority of these MBHs are found (or looked for) in dwarf and bulgeless galaxies. Due to the peculiar hosts and the relatively low BH masses, the non-accretion based methods may start to fail below $\sim10^5$~M$_{\odot}$ \citep[e.g.][]{Reines:2013ys}. On the other hand, a few of the MBHs were studied in the radio band \citep[e.g.][]{Wrobel:2006tg,Wrobel:2008hc,Reines:2011uq,Nyland:2012zr,Secrest:2013fv,Reines:2014ly,Paragi:2014uq} and in all cases the radio emission is likely due to BH activity.

Observable IMBHs have been proposed \citep[e.g.][]{Kaaret:2001ya,Miller:2004kx} in the context of ultraluminous X-ray sources (ULXs; $L_X>3\times10^{39}$~erg~s$^{-1}$) as well as in globular clusters (GCs) \citep[e.g.][and references therein]{Miller:2002rq,Maccarone:2004cq}. ULXs are non-nuclear X-ray sources with a luminosity exceeding the Eddington-limit of a $20-M_{\odot}$ BH, residing in external galaxies. However, most of the ULX population might be explained by mechanical beaming \citep[e.g.][]{King:2009ys}, super-Eddington accretion \citep[e.g.][]{Begelman:2002fb,Poutanen:2007vn}, and massive stellar-mass BHs \citep[e.g.][]{Zampieri:2009if,Belczynski:2010kx}. Relativistically beamed X-rays has also been proposed \citep{Kording:2002xz}, that is difficult to reconcile with the X-ray spectral behaviour of ULXs (see later). Whether there is a link between bulgeless galaxies, dwarf galaxies, and ULXs is still an open question \citep[e.g.][]{Somers:2013ve,Prestwich:2013bh,Plotkin:2014fk}.

A sub-class of ULXs, called Hyperluminous X-ray sources (HLXs), have X-ray luminosities of $L_X \geq 10^{41}$~erg~s$^{-1}$. Their luminosities deem them to be the best candidate IMBHs among ULXs as the above mechanisms can not coherently explain such a powerful output from a stellar-mass BH. However, so far only a handful of HLXs have been studied \citep[e.g.][]{Kaaret:2001ya,Gao:2003fk,Farrell:2009ys,Sutton:2012kx}. The record-holder is HLX-1 in ESO 243-49 with its peak X-ray luminosity of $L_X\simeq10^{42}$~erg/s with a BH mass of $M_{\rm{BH}}\simeq(0.3-30) \times 10^4$~M$_{\odot}$ \citep{Davis:2011oq,Godet:2012ys,Webb:2012kx,Straub:2014fk}. HLX-1 was proposed to be the core of a tidally stripped dwarf \citep{Farrell:2012vn,Farrell:2014cr} or a bulgeless satellite galaxy \citep{Mapelli:2013dq}. \citet{Lasota:2011fk} proposed that HLX-1 may be a binary system with an eccentric orbit. In this case, the BH accretes material from a donor star when it passes periastron, which is thought to drive HLX-1 into regular X-ray outbursts. A wind fed binary system might also be a viable explanation \citep{Miller:2014bs}.
 
 The X-ray behaviour of HLX-1 is remarkably similar to BH X-ray binaries (BHXRBs) \citep[e.g.][]{McClintock:2006fv,Fender:2009bh}. It shows X-ray state transitions \citep{Godet:2009yj} and follows the standard disk law between disk luminosity and temperature ($L\propto T^4$) in its soft state \citep{Servillat:2011fd,Godet:2012ys}. In contrast, the majority of the ULXs may be in a distinct  or ``ultraluminous" X-ray state \citep[e.g.][]{Feng:2009wq,King:2009ys,Gladstone:2009fk}, however, see also \citet{Miller:2013fk} for a number of ULXs that are broadly consistent with the standard disk law. That said, there is one other HLX, M82 X-1, that also follows the trend of $L\propto T^4$ \citep{Feng:2010fk}.

Radio sources offer the best testbed for the scale invariance of the jets under BH mass and/or accretion rate \citep{Heinz:2003fk} and for the disk-jet coupling \citep[e.g.][]{Falcke:1995yb,Merloni:2003ri,Falcke:2004wm}. Although, radio emission as a signature of an accreting IMBH in GCs is yet to be discovered \citep[e.g.][and references therein]{Cseh:2010vn,Wrobel:2011qf}. {On the other hand,} \citet{Webb:2012kx} found that HLX-1 exhibits radio flares (or jet ejections) when the source transits from a hard to soft X-ray state, similar to the behaviour of BHXRBs. \citet{King:2013ij} also found similarities to BHXRBs in NGC 4395 in a sense that its disk-jet coupling could belong to the so-called outliers \citep[e.g.][]{Coriat:2011qm} on the radio-X-ray luminosity plane.
 
In this paper we present radio observations of HLX-1 during its X-ray hard state with the Australia Telescope Compact Array (ATCA) as well as follow-up observations of its flaring emission during its outburst with the Karl G. Jansky Very Large Array (VLA). Throughout this paper, we use a spectroscopically measured distance of HLX-1 of 95~Mpc \citep{Wiersema:2010zr}.

\begin{table*}
\caption{Summary of observations \label{obs}}
\begin{tabular}{ccllclcc}
\hline\hline
Array&Array&Observation&On-source&Central&Reached&Flux&HLX-1 outburst\\
&configuration&date&time&frequency&rms noise& density&status\\
\hline
VLA&A&2011 Aug 23&1 hr&5.9 GHz&7 $\mu$Jy/b&$\leq$21 $\mu$Jy&during burst\\
\hline
ATCA&6D&2012 May  26&7.5 hr&6.7 GHz&7 $\mu$Jy/b&24$\pm$7 $\mu$Jy&hard state\\
ATCA&6D&2012 May  27&7 hr&6.9 GHz&8 $\mu$Jy/b&29$\pm$8 $\mu$Jy&hard state\\
ATCA&6D&2012 May  28&7 hr&6.9 GHz&7 $\mu$Jy/b&$\leq$21 $\mu$Jy&hard state\\
ATCA&6A&2012 Aug 19&9 hr&6.7 GHz&7 $\mu$Jy/b&25$\pm$7 $\mu$Jy&hard state\\
Combined&--&May-Aug&30.5 hr&6.8 GHz& 3.3 $\mu$Jy/b&22$\pm$5 $\mu$Jy&hard state\\
Combined&--&May-Aug&30.5 hr&5.5 GHz& 3.7 $\mu$Jy/b&23$\pm$5 $\mu$Jy&hard state\\
Combined&--&May-Aug&30.5 hr&9.0 GHz& 5.8 $\mu$Jy/b&19$\pm$8 $\mu$Jy&hard state\\
\hline
VLA&B&2012 Sep 1&1 hr&9 GHz&6 $\mu$Jy/b&$\leq$18 $\mu$Jy&during burst\\
VLA&B&2012 Sep 3&1 hr&9 GHz&6 $\mu$Jy/b&$\leq$18 $\mu$Jy&during burst\\
VLA&B--$>$BnA&2012 Sep 5&1 hr&9 GHz&6 $\mu$Jy/b&$\leq$18 $\mu$Jy&during burst\\
VLA&BnA&2012 Sep 23&1 hr&5.5 GHz&6 $\mu$Jy/b&$\leq$18 $\mu$Jy&burst decay\\
VLA&BnA&2012 Sep 24&1 hr&5.5 GHz &7 $\mu$Jy/b&$\leq$21 $\mu$Jy&burst decay\\
VLA&BnA--$>$A&2012 Sep 25&1 hr&5.5 GHz&7 $\mu$Jy/b&$\leq$21 $\mu$Jy&burst decay\\
VLA&A&2012 Oct 1&40 min&5.5 GHz&9 $\mu$Jy/b&$\leq$27 $\mu$Jy&burst decay\\
Combined VLA&--&2011-2012&7.66 h&7.25 GHz&2.1 $\mu$Jy/b&$\leq$6.3 $\mu$Jy&during burst\\
Combined VLA, 60 k$\lambda$&--&2011-2012&7.66 h&7.25 GHz&3.1 $\mu$Jy/b&$\leq$9.3 $\mu$Jy&during burst
\end{tabular}
\medskip

The table shows the individual and combined observations of HLX-1 carried out with the VLA and the ATCA, the array configuration, and the date of the observations. We list the reached rms noise levels at the corresponding central frequency. The table also shows the flux densities obtained from modelling the visibilities or the 3-$\sigma$ upper limits that correspond to a point source. The last column reflects the status of the outburst of HLX-1. The point source flux densities were estimated from the peak brightnesses for the cases of individual detections.
\end{table*}

\section{Observation, Analysis, and Results}

\begin{figure*}
\rotatebox{0}{\resizebox{7.2in}{!}{
\includegraphics{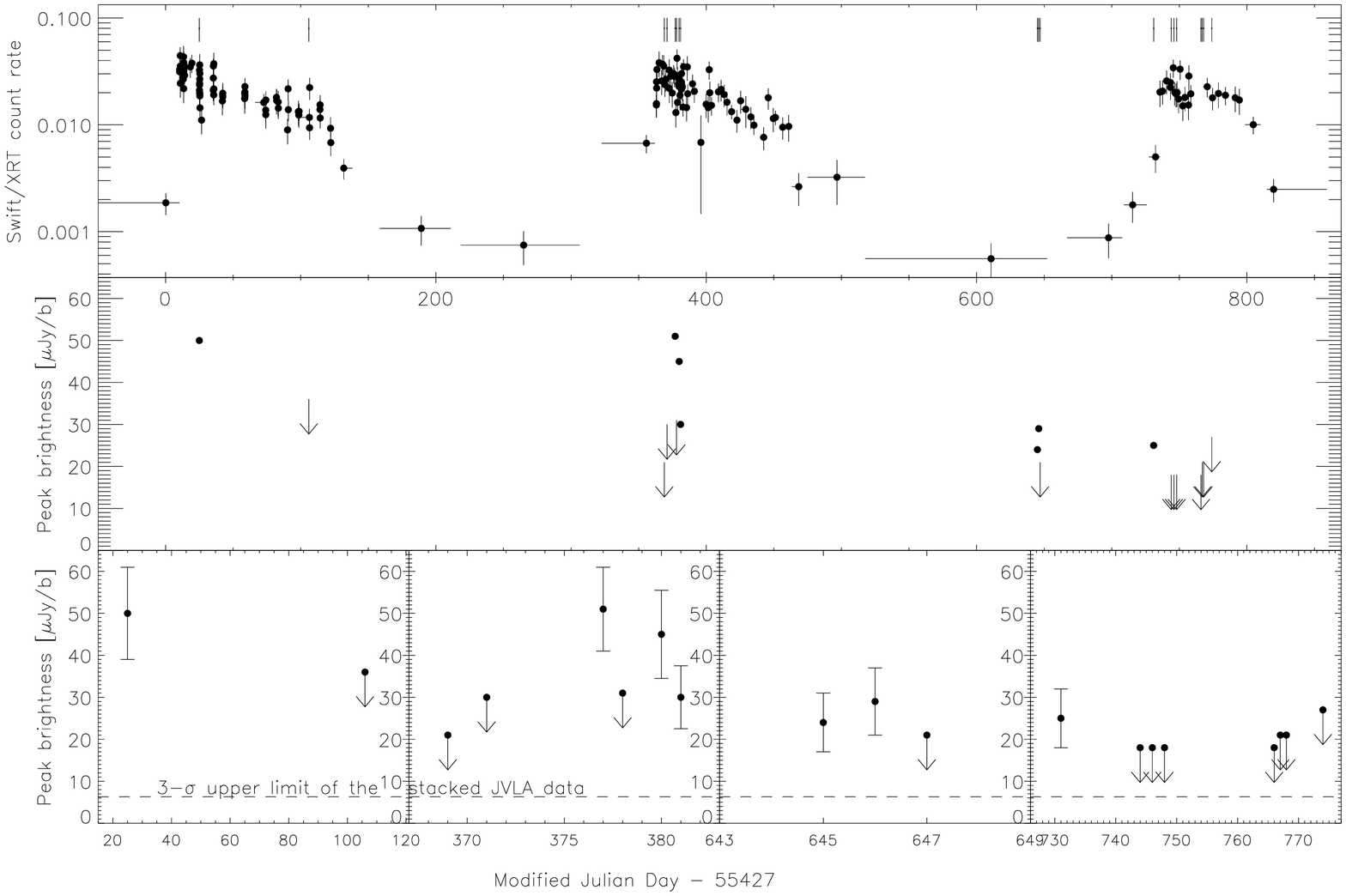} } }
\caption{The figure shows the evolution of the  2010, 2011, and 2012 outburst of HLX-1 along with radio measurements. {\bf The top} panel shows the {\it Swift} X-ray light curve with count rates in the $0.3-10$~keV band. The ticks above each outburst indicate the epochs of radio observations. {\bf The middle} panel shows the corresponding radio measurements. Detections are indicated with dots and non-detections are shown with an arrow. The non-detection values correspond to 3-$\sigma$ upper limits using the rms noise level. {\bf The bottom} panel also shows the same radio data, but with zoomed subsets for clarity. The first subset correspond to the 2010 outburst, the second subset to the 2011 outburst. The third subset shows observations from May 2012, while the fourth subset shows the data corresponding to the 2012 outburst. The bottom panel also shows the deepest 3-$\sigma$ VLA upper limit. Radio points are drawn from Table \ref{obs} and from \citet{Webb:2012kx}.}
\label{lc}
\end{figure*} 

\subsection{{\it Swift} X-ray light curve}

The X-ray light curve of HLX-1 is shown in Fig. \ref{lc}, indicating the evolution of the  2010, 2011, and 2012 outburst in the 0.3-10 keV band together with radio measurements of HLX-1 of this paper and from \citet{Webb:2012kx}. The X-ray light curve was adapted from \citet{Webb:2014fk} and for data reduction and analysis we refer to \citet{Godet:2012ys,Webb:2014uq}. HLX-1 has a fast rise exponential decay (FRED) type of X-ray light curve and the source rapidly transits from a hard to soft state within days, right after the onset of the outburst. After reaching the peak X-ray luminosity, it begins a decay to quiescence over a timescale of months with decreasing duration \citep[e.g.][]{Godet:2014fk}.

\subsection{ATCA observations}\label{atcaob}

We obtained radio observations to look for further similarities between HLX-1 and the phenomenology of BHXRBs, that are known to show flat-spectrum radio cores during their X-ray hard state and are known to show radio flares which are associated with their X-ray state transitions \citep[e.g.][]{Fender:2009bh}. The ATCA measurements were obtained during the hard state of HLX-1 \citep{Godet:2012ys,Webb:2014uq}. So, we combined all of our ATCA data from May and Aug 2012 (see Sec. \ref{atcaob}). The rest of the radio measurements, shown in Fig. \ref{lc}, were obtained during one of the X-ray outbursts of HLX-1 when the source was in a soft state \citep{Servillat:2011fd,Godet:2012ys}, following an X-ray state transition. We obtained the VLA data during the 2011 and 2012 outburst to catch potential flaring events, and then we combined these observations according to the X-ray state of HLX-1 (see Sec. \ref{vlaob}). 

\begin{figure*}
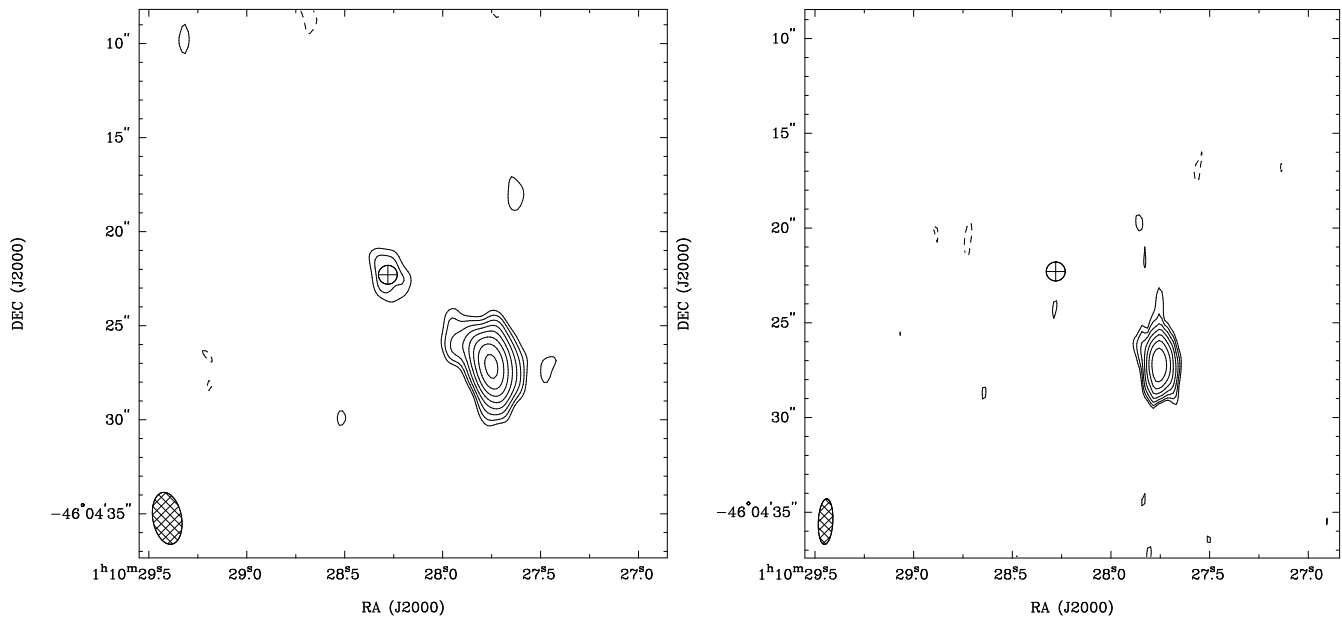

\rotatebox{270}{\resizebox{3.2in}{!}{
\includegraphics{Fig2.ps}}}
\rotatebox{270}{\resizebox{3.2in}{!}{
\includegraphics{Fig3.ps}}}
\caption{{\bf Left:} The combined ATCA image of HLX-1 during the X-ray hard state at 6.8~GHz central frequency. The circle with a cross shows the 95\% positional uncertainty of the Chandra \& HST counterpart of HLX-1 with a radius of 0.5" \citep{Farrell:2012vn}. The radio image was made using natural weighting. The contours are drawn at $9.9 \times (\sqrt{2})^{n} \mu$Jy beam$^{-1}$. The lowest contours represent radio emission at $\pm3$ times the rms noise level of 3.3~$\mu$Jy beam$^{-1}$. The peak brightness is 119~$\mu$Jy~beam$^{-1}$. The Gaussian restoring beam size is $2.79" \times 1.53"$ at a major axis position angle of 10$^\circ$. {\bf Right:} The combined VLA image of the region of HLX-1 showing its host galaxy, ESO~243-49, at 7.25~GHz central frequency. The circle with a cross shows the 95\% positional uncertainty of the Chandra \& HST counterpart of HLX-1 with a radius of 0.5". The radio image was made using natural weighting. The contours are drawn at $6.3 \times (\sqrt{2})^{n} \mu$Jy beam$^{-1}$. The lowest contours represent radio emission at $\pm3$ times the rms noise level of 2.1~$\mu$Jy beam$^{-1}$. The peak brightness is 69~$\mu$Jy~beam$^{-1}$. The Gaussian restoring beam size is $2.40" \times 0.76"$ at a major axis position angle of --3$^\circ$.}
\label{HS}
\end{figure*}

We observed HLX-1 at a location of 1$^h$10$^m$28$^s$.28, $-46^{\circ}$~4'~22".3 \citep{Webb:2010dq,Farrell:2012vn} with the CABB-upgraded \citep{Wilson:2011lg} ATCA in configuration 6D and 6A (baselines up to 6 km) in 2012 under program code C2361 (PI Webb). The data were taken using the CFB 1M-0.5k correlator setup with 2 GHz bandwidth and 2048 channels, each with 1 MHz resolution; and with central frequencies of 5.5 GHz and 9 GHz simultaneously. The on-source integration times were 7-9~hr, see Table~\ref{obs} for more details. The primary amplitude calibrator was PKS~1934-638 and the phase calibrator was 0048-427. The primary was observed for 10 min and the secondary was observed every 15 min. The data reduction was performed using the Miriad software package \citep{Sault:1995lo}. 

First, we combined the 5.5 GHz and 9 GHz visibility data to enhance sensitivity at each epoch. Then, we further combined the observations from May and Aug 2012 as these observations were carried out in the low/hard state of HLX-1 to reach the best sensitivity. Furthermore, we made a separate combination of these data in each frequency band. To produce images we used the multi-frequency synthesis (MFS) method \citep{Sault:1999zr} and the measured system temperatures to compute the weight of the visibility points by the inverse noise variance (instead of integration time). The images were made without self-calibration and with natural weighting. 

The resulting rms noise levels are listed in Table~\ref{obs}. We detected HLX-1 on 26, 27 May and on 19 Aug with a peak brightness of $24\pm7$, $29\pm8$, and $25\pm7$~$\mu$Jy~beam$^{-1}$, respectively. (The 3-$\sigma$ upper limit of 21~$\mu$Jy~beam$^{-1}$ on 28 May is consistent with the above values within the errors). Due to the low signal to noise ratios, the point source flux densities, listed in Table~\ref{obs}, were estimated from the peak brightnesses for the cases of individual detections. The resulting image from the combined data of May and Aug 2012 is shown in Fig. \ref{HS}. In this deepest ATCA image, we detected emission at the location of HLX-1 during the hard state with a signal to noise ratio of $\sim$5.5 with a peak brightness of $18\pm3.3$~$\mu$Jy~beam$^{-1}$. We fitted both, a point-source model and a Gaussian that resulted in $S_{\nu,\rm{Point}}=22\pm3.5$~$\mu$Jy and $S_{\nu,\rm{Gauss}}=26.5\pm5.5$~$\mu$Jy that are consistent within the errors. A point-source model adequately describes the radio emission. However, without higher resolution observations it is not possible to distinguish between a point source or an extended source. We note that the combination of the ATCA data and the VLA data or the combination of the hard-state data with non-hard state data are beyond the scope of this paper. 

To test the fitted values and to minimise the error we attempted to fit the data on the visibility plane. To this purpose we used the Miriad task {\sc uvfit}. First, we attempted to fit the individual epochs, however, due to the low signal to noise ratio the fit did not converge. Regarding the combined data in the individual bands, we fit a point source to HLX-1 and a Gaussian to the host galaxy simultaneously and found $S_{\nu}$=23$\pm$5 $\mu$Jy at 5.5~GHz and $S_{\nu}$=19$\pm$8 $\mu$Jy at 9~GHz. For the deepest ATCA data at 6.8 GHz, we found $S_{\nu}$=22$\pm$5 $\mu$Jy for HLX-1.

\subsection{VLA observations}\label{vlaob}

We observed HLX-1 using the VLA in 2011 and 2012 under project code 11A-290 (PI Krauss) and 12B-024 (PI Cseh), with filler scheduling priority. The observations were carried out in B, A, or hybrid array configuration in X or C band, see Table \ref{obs} for more details. The typical total on-source time was 1~hour that is limited by the low declination of the target. The total bandwidth was 2~GHz, covering either 8--10~GHz (X band) or 4.5--6.5~GHz (C band), that consists of two base bands with 8-bit sampling, each with 8 sub-bands with 4 polarisation products, 64 channels per sub-band with 2~MHz channel width. In 2011, the lower and the upper baseband was placed as 4.5--5.5 and 6.2 --7.2~GHz.

The data were calibrated and imaged in CASA 4.1.0  \citep{McMullin:2007ys} and we applied the VLA calibration pipeline version 1.2.0. \footnote{https://science.nrao.edu/facilities/vla/data-processing/pipeline} We used 3C48 for absolute flux density and bandpass calibration, and J0106-4034 was used for antenna gain and phase calibration. The primary calibrator was observed for 10 min and the phase calibrator for 2 min out of every 10 min. Images were made at each epoch using the MFS method and natural weighting. There was no self-calibration applied. The pipeline calibration products and the image quality -- i.e. the reached rms noise levels and noise structures -- were then visually examined. We also compared these to a set of manually calibrated data, and found equally good or better image quality of the pipeline calibrated data due to a more robust flagging strategy.

Our primary aim was to detect radio flares, study their time evolution, and to test if there is a permanent emission, like a ULX radio bubble \cite[e.g.][]{Cseh:2012fk}, at low flux density levels. The individual runs are in principle suitable to look for flares from HLX-1. However, we did not detect emission at or above the 3-$\sigma$ rms level (that corresponds to the individual epochs, see Table~\ref{obs}) at the location of HLX-1 in these observations.

Furthermore, to test the presence of a permanent or an extended emission, all of the VLA visibility data were combined and then imaged to reach the best rms noise level so far. This is shown in Fig. \ref{HS} right, where no radio emission is detected at the location of HLX-1 above a 3-$\sigma$ upper limit of $\sim6.3~\mu$Jy. To test whether we could have resolved out large-scale, extended structures around HLX-1, we again combined all the visibility data and we restricted the maximum $(u,v)$ range to 60~k$\lambda$. This reduction of the apparent maximum baseline length produces a much larger synthesised beam width allowing for the detection of emission from a larger spatial scale. In practice, the largest angular scale (LAS) that can be imaged comes from the shortest baselines and for these array configurations and observing bands the LAS is 4.5-8.5 arcsec\footnote{https://science.nrao.edu/facilities/vla/docs/manuals/oss/\\performance/resolution}.  For our combined data, we take a conservative LAS of $\sim4.5$ arcsec that corresponds to A configuration in C band. The resulting synthesised beam width is 4.70"$\times$2.40", within the LAS, and the resulting rms noise level is 3.1 $\mu$Jy/beam. We do not detect radio emission at the location of HLX-1. The VLA data supports that there is no permanent or extended emission present smaller than a physical scale of 2.2 kpc $\times$ 1.1 kpc.

Here we also provide the peak brightness and the total flux density of the host galaxy of HLX-1 that is fitted with a Gaussian and the errors given are statistical. These values correspond to Fig. \ref{HS} left, right, and to the $(u,v)$-restricted VLA image. Furthermore, given the low elevation of our target, we attempt to quantify the potential atmospheric attenuation in X band. To this end, we compare the flux density and the fitted position of the host galaxy between the ATCA and the VLA at 9~GHz central frequency. We combined all of the VLA data observed in X band and restricted the $(u,v)$ range to 60~k$\lambda$ to minimise the flux density loss due to the higher resolution of the VLA. Comparing the flux densities, we find that they agree within the statistical errors (Table \ref{host}). However, given the relatively high uncertainties on the fitted flux densities and the lack of a more suitable source on the target field, we quote a conservative atmospheric attenuation of $<24$\% by taking the flux densities at face value and assuming a non-variable source. We also provide flux densities for the ATCA data, obtained from fitting a Gaussian on the visibility plane, that may indicate a flux density overestimate for the image-plane Gaussian fits and thus less of a difference at high frequencies. The fitted positions differ by only 0.02 arcsec in RA and 0.14 arcsec in Dec that is less than our cell size.

\begin{table}
\caption{Host galaxy radio properties}\label{host}
\begin{tabular}{cccccc}
\hline\hline
Array&Peak&Flux density&Beam&Freq.\\
&[$\mu$Jy/b]&[$\mu$Jy]&[" $\times$ "]&[GHz]\\
\hline
ATCA&119$\pm$9&157$\pm$16 (153$\pm$6)&2.79 $\times$1.53&6.8\\
VLA&68$\pm$4&116$\pm$9&2.40 $\times$0.76&7.25\\
VLA, 60k$\lambda$&100$\pm$7&122$\pm$12&4.70 $\times$2.40&7.25\\
ATCA&117$\pm$12&167$\pm$24 (143$\pm$11)&2.30 $\times$1.23&9.0\\
VLA, 60k$\lambda$&91$\pm$9&127$\pm$19&5.99 $\times$2.52&9.0\\
\end{tabular}
\medskip

The table shows the image-plane Gaussian fit results of the host galaxy. Flux density values in brackets indicate a Gaussian model fit on the visibility plane.
\end{table}

\subsection{Spectral index and variability}

We attempted to estimate a spectral index of the radio source detected with ATCA from the combined images in the different bands and found a two-point spectral index between 5.5 and 9 GHz of $\alpha=-0.4\pm1.0$, where $S_{\nu}\sim\nu^{\alpha}$. This spectral index does not constrain the radio spectrum to be steep, flat, or inverted.

We aimed to repeat this exercise on the previous individual detections in 2010 and 2011, using peak brightnesses in the different bands; see the values in the supplementary material of \citet{Webb:2012kx}. However, the frequency-split data do not show detections on 4 Sep 2011 due to the high noise level. Furthermore, in the remaining epochs, detections are present in only one of the frequency bands. Therefore one can set upper or lower limits to the spectral indices using 3-$\sigma$ upper limits from the counter band. We find spectral indices of $\alpha_1<-0.5\pm0.5$ on 2010 Sep 3, $\alpha_2<0.3\pm0.5$ on 2011 Aug 31, and $\alpha_3>1.6\pm0.4$ on 2011 Sep 3. The spectral index of $\alpha_1$ indicates a flat to steep spectrum, $\alpha_2$ does not constrain the radio spectrum, and $\alpha_3$ indicates an inverted radio spectrum. \citet{Webb:2012kx} showed that the radio emission in 2010 and 2011 was variable. Considering our deepest VLA limit and comparing it to the previously measured peak brightnesses, we find that the emission is variable by a factor of about ten, which unambiguously confirms a variable radio source in 2010 and 2011. The apparent variation in the spectral indices ($\alpha_1,\alpha_2,\alpha_3$) is also expected for a variable jet source. 

Regarding the ATCA data in 2012, the peak brightness values are constant within the errors. Albeit, the non-detection on 28 May might indicate variability that is difficult to assess due to the high rms noise levels. However, in the deepest VLA limit the probed spatial scales are coupled with a sensitivity that is better than and matched to that of the ATCA, which argues against a permanent radio source in 2012.

\section{Discussion}

In the following, we discuss the radio emission detected at the location of HLX-1 during the hard state. First, we investigate whether this emission could be contaminated by permanent emission such as a radio bubble or star formation. Then, we discuss potential origins of the emission due to a compact jet or an interacting jet with the ISM. Furthermore, we discuss the properties of the flaring emission and compare with microquasars and active galactic nuclei (AGNs).

\subsection{Permanent emission: radio bubble or star formation?}\label{bubble}

The smallest resolvable angular scale is 0.76 arcsec (Fig. \ref{HS} right), that corresponds to a physical scale of $\sim$350~pc at 95~Mpc. Some of the largest microquasar and ULX radio bubbles, like S26 and IC 342 X-1 \citep{Soria:2010lt,Cseh:2012fk}, have a size of $\sim200-300$~pc and a high radio luminosity of $L_R\simeq2 \times 10^{35}$~erg/s that would correspond to $\sim$3~$\mu$Jy at a distance of HLX-1. These sources would not be resolved and would not reach our detection threshold. On the other hand, there is at least one peculiar example of an extended radio source, assumed to be associated with the ULX NGC 2276 3c, with a size of 650~pc and with an integrated flux density of 1.4~mJy at 33.3~Mpc \citep{Mezcua:2013uq} that would be detectable as well as resolvable if placed at 95~Mpc. This source was also inferred to be powered by an IMBH, that may allow a better comparison with HLX-1. Given the that one would have detected ample extended emission if similar to NGC~2276~3c, HLX-1 is unlikely to be associated with any radio bubble above a flux density level of 6.3~$\mu$Jy.

\citet{Farrell:2012vn} argued for the presence of a young star cluster hosting HLX-1. In the following we estimate the expected radio flux density due to star formation. We take a near UV (NUV) flux of 23.96$\pm$0.04 mag at wavelength of 2829.8 \AA \, that is obtained from {\it HST} photometry \citep{Farrell:2012vn}, which corresponds to a NUV luminosity of $L_{\rm{NUV}}=(1.08\pm0.04) \times 10^{40}$~erg/s. Assuming that all of this emission is stellar in origin, we estimate the star formation rate (SFR) using the relation of  $\log \dot{M_{\star}}/(\rm{M_{\odot}\, yr^{-1}}) = \log L_{\rm{NUV}}/(\rm{erg\, s^{-1}}) - 43.17$ \citep{Kennicutt:2012kx}. Given that the accretion disk is expected to significantly contribute to the NUV emission, we place an upper limit to the SFR and find that $\dot{M_{\star}}\leq7.3 \times 10^{-4}$~M$_{\odot}$~yr$^{-1}$.

Using our SFR estimate, the expected thermal and non-thermal radio flux density can be obtained \citep{Murphy:2011vn,Nyland:2012zr} from
\begin{equation}
\begin{split}
& \left(\frac{L_{\nu}}{\rm{erg\,s^{-1} Hz^{-1}}}\right)=10^{27}\left(\frac{\dot{M_{\star}}}{\rm{M_{\odot}\, yr^{-1}}}\right) \times \\
& \left[2.18\left(\frac{T_e}{\rm{10^4\,K}}\right)^{0.45} \left(\frac{\nu}{\rm{GHz}}\right)^{-0.1}+15.1\left(\frac{\nu}{\rm{GHz}}\right)^{\alpha_{\rm{NT}}}\right],
\end{split}
\end{equation}
where we take $T_e=10^4$~K, $\alpha_{\rm{NT}}=-0.8$ and $\nu=7$~GHz. We find that the total radio flux density at 7~GHz due to SF would be $L_{\nu}\leq3.64 \times 10^{24}$~erg~s$^{-1}$~Hz$^{-1}$ that corresponds to a flux density of $\leq$0.34~$\mu$Jy at a distance of HLX-1. So, the ATCA radio emission is unlikely to be due to star formation or a radio bubble.

\subsection{Mass, accretion rate, and jet power}\label{jet}

We detected radio emission during the hard state with a total flux density of $S_{\nu}=22\pm3.5$~$\mu$Jy at a central frequency of 6.8~GHz at a significance of 5.5~$\sigma$. A point source model adequately describes the radio emission and it is unlikely to be permanent. Given that ``compact jets" are ubiquitous in a hard or inefficient accretion state and have a self-absorbed (flat or slightly inverted) radio spectrum, the radio emission detected at the location of HLX-1 can be consistent with a compact jet. Alternatively, this radio emission may not be related to the X-ray state of HLX-1 and it could perhaps be a residual emission due to the flaring activity of the source in a form that the ejected components could interact with the large-scale ISM. We discuss this option in the following sections. Here, we assume that the radio emission is causally connected to the X-ray hard state of HLX-1 and it is a compact jet as seen in other BHXRBs in this state.

Under this assumption, in the following we employ the ``fundamental plane" of BH activity \citep{Merloni:2003ri,Falcke:2004wm,Kording:2006pt,Plotkin:2012kx} to estimate the BH mass and use the flux density of the compact jet to estimate the accretion rate in the hard state, and total jet power. The correlation was found to be valid for hard X-ray state objects, where the accretion is jet-- or advection dominated, and with a small intrinsic scatter of 0.07 dex \citep{Plotkin:2012kx}. To estimate the BH mass we use the fit of \citet{Miller-Jones:2012uq}, who re-regressed the plane with the mass as the independent variable:
\begin{equation}
\log M_{\rm{BH}} = 1.638\log L_R - 1.136\log L_X -6.863.
\end{equation}
Here, $L_X$ is the X-ray luminosity in the 0.5-10 keV band and $L_R$ is the radio luminosity at 5~GHz (L=$\nu$L$_\nu$) in units of erg~s$^{-1}$ and the mass is in ${\rm M_{\odot}}$. The 1$\sigma$ uncertainty on the mass estimate is 0.44 dex. After substituting a typical hard state luminosity of HLX-1 of $L_X=(2.0\pm0.6)\times10^{40}$~erg~s$^{-1}$ \citep{Godet:2012ys} and $L_R=(1.2\pm0.3)\times10^{36}$~erg~s$^{-1}$, the BH mass would be $M_{\rm{BH}}=2.8^{+7.5}_{-2.1} \times 10^{6}$~M$_{\odot}$ at 90\% confidence. This is at least an order of magnitude higher than previous estimates (see later).

Similarly, under the assumption that we witness a compact jet, we derive a total jet power and an accretion rate using Eq. 1 and Eq. 2 of \citet{Kording:2008nq}. We obtain $Q_j\simeq7.2\times10^{36}[L_R/(10^{30}~\rm{erg/s})]^{12/17}\simeq1.4\times10^{41}$~erg/s and $\dot{M}\simeq4\times10^{17}[L_R/(10^{30}~\rm{erg/s})]^{12/17}\simeq7.8\times10^{21}$~g/s (or $1.2 \times10^{-4}$~M$_{\odot}$/yr). 

Considering the jet power, we recall that the integrated X-ray energy of the 2011 outburst is $5.77 \times 10^{48}$~erg \citep{Miller:2014bs}. Averaging over the outburst decay time of 110 days, the average X-ray luminosity is $\bar{L}_{x}=6 \times 10^{41}$~erg/s. On average, a typical value of $\sim$10\% of the accretion power thought to go into to the jets during radio loud phases of accreting black holes \citep[e.g.][]{Falcke:1995yb}. Therefore, the expected average total jet power of HLX-1 is $\bar{Q}_{j}\simeq6\times10^{40}$~erg/s. This average jet power is somewhat lower than the estimate from a hard state compact jet, where \citet{Kording:2008nq} also assumed a 10\% jet efficiency, but consistent within the errors.

\subsubsection{Higher BH mass or an outlier on the fundamental plane?}

After applying the slim disk model of \citet{Kawaguchi:2003uq} to the X-ray spectra of HLX-1, \citet{Godet:2012ys} found a best fit BH mass of $M_{\rm{BH}}\sim2 \times 10^4$~M$_{\odot}$. Similarly, a more sophisticated modelling that employs a relativistic slim disk with full radiative transfer resulted in a mass of $M_{\rm{BH}}\simeq(0.6-19.2) \times 10^4$~M$_{\odot}$, depending on spin and inclination \citep{Straub:2014fk}. On the other hand, the chosen disk model does not affect significantly the BH mass estimate and if applying a standard disk model, the upper end of the mass of HLX-1 was found to be $3\times10^5$~M$_{\odot}$ at 90\% confidence \citep{Davis:2011oq}. On the contrary, the BH mass obtained from the fundamental plane is at least a factor of 10 higher than obtained from the X-ray spectral fits and it is not consistent within the 90 \% errors.

In the model of \citet{Godet:2012ys}, the peak accretion rate is $\dot{M}=4 \times10^{-4}$~M$_{\odot}$/yr and during a hard state it is $\dot{M}<4 \times10^{-6}$~M$_{\odot}$/yr. Also, \citet{Straub:2014fk} found an accretion rate of $\dot{M}=(0.3-2) \times10^{-4}$~M$_{\odot}$/yr during burst. The mass accretion rate deduced assuming the presence of a compact jet is also at least an order of magnitude higher than the accretion rate derived from X-ray spectra and the estimates are not consistent within the errors.

Furthermore, the star cluster that hosts HLX-1, has an inferred mass of $10^5-10^6$~M$_{\odot}$ \citep{Farrell:2014cr} and to have a BH mass to cluster mass ratio in the order of unity is unlikely \citep[e.g.][]{Miller:2002rq}, however, recent results show that in compact dwarf galaxies the ratio could be as high as 15\% \citep{Seth:2014vn}. We assume that the correct BH mass is estimated from the X-ray spectral fits. The resulting high BH mass (and accretion rate) obtained from the fundamental plane could then indicate that HLX-1 is an outlier source. It is either radio bright or X-ray faint with respect to the correlation. A comprehensive study of a sample of such outliers \citep{de-Gasperin:2011uq} concluded that those sources are X-ray faint rather than radio bright. On the other hand, in the case of HLX-1, the hard state X-ray luminosity is only a factor of $\sim100$ lower than the peak X-ray luminosity and therefore we argue that HLX-1 is rather radio bright. So, the radio emission could be Doppler-boosted and the above estimates should then be considered as upper limits. 

\citet{Miller-Jones:2006uq} argues that an unconfined, hard state compact jet may have a Lorentz-factor above $\Gamma\simeq2$. The Lorentz-factor is $\Gamma = (1-\beta^{2})^{-0.5}$, where $\beta$ is in units of speed of light and the Doppler-factor is $\delta=\left[\Gamma(1-\beta \cos \phi)\right]^{-1}$, where $\phi$ is the viewing angle. The Doppler enhancement is proportional to $\delta^{3-\alpha}$ and it has a maximum at $\delta_{max}\simeq2\Gamma$. Here, we conservatively assume that our compact jet has a $\Gamma=1.4-2$. Then, the radio emission could be boosted by a factor of up to $22-64$ for a flat-spectrum jet\footnote{Applying this value to the radio emission and using the fundamental plane would still give a mass estimate consistent with the X-ray spectral fits.}. To have a consistent mass estimate from the fundamental plane, the radio emission of the compact jet has to be boosted by at least a factor of $\sim5$, which would limit the viewing angle to $\theta \leq 34^{\circ}$ (to the angle where $\delta^3 \geq 5$). On the other hand, in Sec. \ref{fl}, based on the energetics we argue that the flares may not be boosted or could even be deboosted at the same time. This could be the case for certain viewing angles if the jet has a higher Lorentz-factor during a flaring event. Assuming a maximum of $\Gamma=10$ for a flare would limit the viewing angle to $\theta \geq 25^{\circ}$ (the angle where $\delta\leq1$). Therefore the case of a mildly boosted compact jet and a deboosted flare is possible for a viewing angle range of $\theta \simeq 25^{\circ} - 34^{\circ}$, which would also limit the minimum Lorentz-factor of the flare to $\Gamma=5.5$, assuming identical viewing angles for the compact jet and the flare. In this scenario the maximum Doppler enhancement would be 7.4-12.3 for the compact jet.

Alternatively, a face-on SS433 type system has been proposed for some ULXs \citep{Fabrika:2001fk} and recently specifically for HLX-1 \citep{King:2014uq}, who argue that the geometric beam cone (or ' funnel') has an opening angle of $1.6^{\circ}$. This is a tempting geometric model to apply to our radio observations. Taking a distance of 5.5~kpc and an integrated flux density of S=0.7 Jy for SS433 \citep[e.g.][and references therein]{Miller-Jones:2008uq}, the source would have an integrated flux density of 2.3~nJy if placed at the distance of HLX-1. To match the measured flux density of HLX-1, a Doppler enhancement of at least $\sim$9400 (or $\delta \geq 21.1$) would be required. Assuming that the radio jet has a viewing angle similar to the quoted geometric beam opening angle (i.e. it fits into the funnel), a Lorentz-factor of $\Gamma \geq 12$ would be required. This would mean a highly relativistic jet (0.996~c) that might be in contrast to SS433 whose jet is likely to be mass-loaded and has a speed of only 0.26~c. In any case, this geometric model could be applied to the radio emission, albeit, this model does not explain the X-ray spectra and state transitions of HLX-1 \citep{King:2014uq}.

\subsection{Flare properties of HLX-1}\label{fl}

In this Section, we investigate the energetics and kinematics of the radio flares detected in 2010 and 2011 and discuss the possibility that the radio emission detected in 2012 is not related to the X-ray hard state.

\subsubsection{Flaring behaviour}

We observed HLX-1 during its outburst with the VLA to catch flares and to study their time evolution. However, we did not detect new flares in our 2012 campaign. With respect to the deepest VLA noise level, we find that the previously detected flares are variable by a factor of ten. Also, the apparent variation of the spectral indices ($\alpha_1,\alpha_2,\alpha_3$) is consistent with a variable jet source. This is because $\alpha_3$ indicates an inverted radio spectrum that is peaking in the 9-GHz band (or higher). Within the errors, it may be more inverted than allowed by a synchrotron self-absorbed core of a conical jet of $\sim1.5$ \citep[e.g.][]{Paragi:2013kx}, that may indicate a flare or discrete ejection. Inverted spectra are typically seen in synchrotron self-absorbed objects, where the signature of the evolution (or age) of an ejection is that the synchrotron turnover frequency shifts from high to low frequencies with time due to the expansion and the radiative losses of the ejected material. However, the exact dynamical evolution of a flare will depend on the ejected mass, on its initial Lorentz-factor, and ISM density \citep[e.g.][]{Peer:2012fk}. There are example radio spectra of flaring microquasars, like GRS1915+105, that show a turnover frequency at $\sim1-3$~GHz \citep{Rodriguez:1995kx}. In the case of Cyg X-3, the turnover is at $\sim7-10$~GHz  \citep[e.g.][]{Marti:1992vn}. This kind of radio spectrum is also exhibited by AGN flares, where the actual shift of the turnover frequency may start from 10s-100s of GHz and could be tracked by having several broadband radio spectra measured, that is allowed by a much longer flaring timescale \citep[e.g.][]{Hovatta:2008fk}. In this regard, HLX-1 exhibits radio properties that are typical for flares in jets that also complies with the interpretation of the radio emission being due to flares \citep{Webb:2012kx}.

Furthermore, the X-ray evolution of HLX-1 shows state transitions \citep{Servillat:2011fd,Godet:2012ys} similar to microquasars \citep[e.g.][]{McClintock:2006fv,Fender:2009bh} and radio emission was detected following these transitions \citep{Webb:2012kx}. These flares are thought to occur after crossing the ``jet line" \citep{Fender:2009bh} on the hardness-intensity diagram of the outbursts, still, it is not straightforward to estimate an exact arrival time. In HLX-1 radio flares were detected after reaching the X-ray peak (Fig \ref{lc}), when the source was already in a disk dominated state, where the radio emission is quenched in most cases of microquasars. On the other hand, as observations show, flares can indeed occur during a purely disk dominated state and some sources do not seem to completely quench in radio \citep{Brocksopp:2002cr,Brocksopp:2007fk,Brocksopp:2013uq}. Nevertheless, in some cases it is also common to have a major flare that is later followed by multiple smaller ones, see e.g. Fig. 4 of \citet{Brocksopp:2002cr}. \citet{Webb:2012kx} showed a detection of at least two flares from HLX-1 in 2011, that may indicate similar flaring activity in HLX-1. However, \citet{Brocksopp:2013uq} have shown that the different subsequent flares have diverse spectra, duration, and morphology, which could be due to multiple ejecta or due to rebrightening. This is because instead of a simple flux decay, as a result of a ballistic motion, the rebrightening may reflect an additional deceleration due to the interaction with an earlier ejecta or with the ISM. 

Here, we account for the possibility that the radio emission detected in 2012 is not connected causally to the X-ray state, rather to the flaring activity. In this case, one has to explain a variable source, as argued in Sec. \ref{bubble}, that slowly decelerates. The average flux density of the flares in 2011 was about 45~$\mu$Jy, that then dropped to $\sim22$~$\mu$Jy within $9-11$~months, considering the detections in 2012 May and Aug. This hard state flux density then dropped below 6.3~$\mu$Jy in 1 month, when the first VLA observations were carried out. In principle, such a slow deceleration might not be excluded as some microquasars show ejecta interaction with the ISM on timescales up to years  \citep{Corbel:2002qf} and we would not be able to spatially resolve such components at the distance of HLX-1. On the other hand, the number of measurements are not sufficient to model a decay time. Still, we posit that a simple exponential flux decay is unlikely on the above timescales. This is because the flare emission detected in 2011 has already varied within a few days \citep{Webb:2012kx}, that may indicate a more rapid cooling. Also, we could easily miss flares in 2012 due to a potentially short flaring timescale, i.e. due to sparse sampling or bad timing, given that the VLA on-source time is limited to 1 hr, while the ATCA integration times were much longer on the order of 7-9 hr. If the flares are variable on the order of a day or shorter, then one has a much lower chance to catch a flare with the VLA.

Flares are typically a factor of 10-100 above a non-flaring continuum \citep{Brocksopp:2002cr,Brocksopp:2013uq} and our $3\sigma$ limit of 6.3~$\mu$Jy may imply that in HLX-1 such continuum level is below this limit or the source does quench between the subsequent flares. Based on this, a compact jet would be expected to have a much lower flux density than the flares, lower than the measured factor of 2-3. Given the relatively short variability timescale and the apparently high flux density with respect to the flares, one may also conclude that the radio emission is boosted. Alternatively, we missed the major flare, which may involve the future possibility of detecting flares closer to mJy levels.

\subsubsection{Energetics and beaming}

In the following we assume equipartition and estimate a minimum jet power ($Q_{min}$) of a single associated ejection event corresponding to the highest flux density on 2011 Aug 31 to compare with our average jet power
deduced in Sec. \ref{jet}. This can be estimated from the following equation \citep[e.g.][]{Longair:1994hs,Miller-Jones:2006uq}, that we parametrized for HLX-1:

\begin{equation}
\begin{split}
Q_{min}=4.67 \times 10^{36} \eta^{4/7} & \left( \frac{t_r}{\rm{day}} \right)^{2/7}  \left( \frac{d}{\rm{Mpc}} \right)^{8/7}  \left( \frac{\nu}{\rm{GHz}} \right)^{2/7} \\  
 \times & \left( \frac{S_{\nu}}{\rm{\mu Jy}} \right)^{4/7} \, \rm{erg/s},
\end{split}
\end{equation}
where $\eta-1$ is the ratio of relativistic proton to electron energy, $\nu$ is the observing frequency, $S_{\nu}$ is the observed flux density, $d$ is distance, and $t_r$ is the rise time of the flare. This equation does not account for a kinetic energy associated with the bulk motion of an $e^{\pm}$ plasma (nor of an $e-p$ plasma). Also, in the above equation the volume of the ejecta is derived from its light crossing time, which is approximated by the rise time of the flare. We do not account for relativistic protons ($\eta=1$) and use a flux density of $\sim$63~$\mu$Jy at 7~GHz central frequency. The rise time is not known, however, based on the variability seen in the 2011 observations \citep{Webb:2012kx}, we assume a fiducial value on the order of a day and we provide $Q_{min}$ as a function of this parameter:

\begin{equation}\label{obsframe}
Q_{min}=1.6 \times 10^{40} \left( \frac{t_r}{\rm{day}} \right)^{2/7} \, \rm{erg/s}.
\end{equation}

$Q_{min}$ is not a strong function of the rise time and the uncertainty in our estimate could be dominated by the unknown beaming effects. Such beaming effects are necessary to invoke because the hard state jet power is a factor 10 higher than the jet power of the brightest flare, if taken at face value. So, either the compact jet is boosted or the flares are deboosted. In the following, we investigate the possible intrinsic power of an apparently deboosted flare. 

To have an insight into the intrinsic jet power, the derived $Q_{min}$ has to be transformed to the rest-frame by scaling it by a factor\footnote{Note that this correction factor is model dependent as the emitting volume is estimated from the rise time that also suffers time dilation. For instance, if the volume can be estimated directly from high-resolution observations, then the correction factor needs to be modified accordingly.} of $\delta^{-(3-\alpha)4/7}$  \citep[e.g.][]{Miller-Jones:2006uq}, where $\delta$ is the Doppler-factor. To estimate the correction factor, we consider a deboosted approaching jet -- neglecting a receding jet -- and an acceptable range of $\Gamma\simeq[1.4;10]$. The lower end of this range reflects one of the lowest $\Gamma$ measured among microquasar ejecta \citep{Yang:2011fk}. While, the upper end was chosen to account for a possibly high $\Gamma$, should the ejecta have not been frustrated \citep{Miller-Jones:2006uq}. Given the unknown viewing angle, this range of Lorentz-factors translates to a minimum and maximum $\delta$ of $[0.1; 20]$. Then, a correction factor of up to 193 could be obtained for $\alpha=-1$. Therefore, a deboosted flare might not be excluded. However, as argued in Sec. \ref{jet}, it is unlikely that the intrinsic jet power of an ejection is larger than the overall average output and the boosted compact jet scenario is favoured.

\section{Conclusions}

We investigated the radio properties of HLX-1 using a set of follow-up ATCA and VLA observations. These involved observations during a hard X-ray state (ATCA) as well as at the beginning and during the decay of the 2012 outburst (VLA). We detected radio emission at the location of HLX-1 when the source was in a hard state. Under the assumption that the emission is coming from a compact jet we provide an independent mass estimate of the BH of $M_{\rm{BH}}\leq2.8^{+7.5}_{-2.1} \times 10^{6}$~M$_{\odot}$ at 90\% confidence. We also investigated the possibility that this radio emission could be due to a radio nebulae, star formation, a residual of the flaring activity. We conclude that the detected emission is not compatible with a nebula or star formation and it is unlikely to be due to a jet interacting with the ISM, because the variability seen on much shorter timescales. We argue that the radio emission is likely to be Doppler-boosted making the mass (and accretion rate) estimate an upper limit. The follow-up VLA observations do not reveal new detections, possibly due to an insufficient sampling of a relatively short rise time of the flares. The VLA data suggests a variable radio source and the flare energetics, based on analogies with microquasars, also suggests a boosted radio source. Further radio monitoring is needed to test the time-dependent behaviour and the compactness of the radio emission, perhaps with the MeerKAT or the SKA.

\section*{Acknowledgments}
We thank the anonymous referee for the quality comments that improved the paper. The National Radio Astronomy Observatory is a facility of the National Science Foundation operated under cooperative agreement by Associated Universities, Inc. The Australia Telescope is funded by the Commonwealth of Australia for operation as a national facility managed by CSIRO. The authors are grateful to N. Gehrels for supporting {\it Swift} observations of HLX-1. This work made use of data supplied by the UK {\it Swift} Science Data Centre at the University of Leicester. EK acknowledges fund from an NWO Vidi grant Nr. 2013/15390/EW. SAF was the recipient of an Australian Research
Council Postdoctoral Fellowship, funded by grant DP110102889. The Centre for All-sky Astrophysics (CAASTRO) is an Australian Research Council Centre of Excellence, funded by grant CE110001020.

\bibliographystyle{mn2e} 

{\small \bibliography{my}}
\label{lastpage}

\end{document}